\providecommand{\reserveinserts}[1]{}
\documentclass[APA,STIX2COL]{WileyNJD-v2}

\usepackage{balance}

\makeatletter
\def\NAT@aysep{,}
\makeatother

\articletype{Article Type}%
\usepackage{subcaption}

\received{26 April 2016}
\revised{6 June 2016}
\accepted{6 June 2016}

\begin{document}

\title{From Corridor Selection to Earthwork: A Multi-Stage Framework for Automated Road Design via Steiner Trees and Convex Optimization}

\author[1]{Paavai Manimaran Vanjeenathammal}

\author[2]{John R.J. Thompson}

\author[3]{Warren Hare}

\author[4]{Yves Lucet}

\authormark{Paavai M. Vanjeenathammal \textsc{et al}}

\address[1,4]{\orgdiv{Department of Computer Science}, \orgname{University of British Columbia, Okanagan Campus}, \orgaddress{\state{Kelowna}, \country{Canada}}}

\address[2]{\orgdiv{Department of Statistics}, \orgname{University of British Columbia, Okanagan Campus}, \orgaddress{\state{Kelowna}, \country{Canada}}}

\address[3]{\orgdiv{Department of Mathematics}, \orgname{University of British Columbia, Okanagan Campus}, \orgaddress{\state{Kelowna}, \country{Canada}}}

\corres{*Yves Lucet. \email{yves.lucet@ubc.ca}}

\presentaddress{UBC Okanagan, ASC 350, 3187 University Way, Kelowna BC V1V 1V7 Canada}

\abstract[Summary]{Designing road networks for wind farms in complex terrain is a challenging task, especially under tight construction budgets. Traditional manual methods are time-consuming and may yield suboptimal results. We propose a structured three-phase optimization framework, TriPhase, to automate and minimize road construction costs from corridor selection to earthwork. Phase one formulates corridor selection as a Steiner minimum tree problem over a terrain-aware graph, incorporating slope and curvature constraints. In phase two, each road segment undergoes horizontal alignment optimization using a bilevel model, where a mixed-integer program evaluates vertical alignment costs. To ensure solver compatibility and performance, the model is reformulated explicitly for Gurobi. The final phase applies a network-wide convex optimization model to refine vertical alignment. Numerical experiments on real-world sites demonstrate up to 14\% cost savings compared to industry-standard manual designs, validating the framework’s effectiveness and practical relevance.}


\maketitle

\section{Introduction}

Wind energy has emerged as a vital source of renewable power, with wind farms increasingly deployed in remote and mountainous regions to harness optimal wind conditions. Among the various civil infrastructure requirements, access roads play a crucial role in enabling turbine installation, facilitating maintenance, and ensuring emergency accessibility. While turbines constitute the largest share of investment, access roads and related civil works can account for up to 20\% of the total project cost, especially in steep terrain where earthwork complexity and geometric constraints increase construction expenses~\cite{ZHAO-09}.

Unlike conventional roads that connect two endpoints, wind farm roads must provide efficient access from a single entry point to multiple turbine locations. This requirement introduces a large and complex design space involving multiple trade-offs between terrain feasibility, earthwork volumes, and geometric alignment. Manual design methods, still widely used in industry software, are labor-intensive and unable to guarantee globally optimal solutions. As wind farm projects scale up, suboptimal networks can lead to significantly higher construction costs, environmental impacts, and logistical challenges~\cite{PEREZ-25}.

To address these challenges, we propose a three-phase computational optimization framework, \textit{TriPhase}, that automates road network design from corridor selection to vertical profile refinement. The model integrates terrain-aware graph optimization, bilevel road alignment modeling, and network-wide vertical alignment refinement into a unified cost-minimization approach.

\section{Past Research in Road Design Optimization}

Road design optimization has been an active area of research for several decades, with significant contributions spanning transportation engineering, operations research and computational optimization. The overarching goal is to produce cost-efficient, safe and environmentally conscious road alignments while adhering to terrain and engineering constraints.

Early efforts in this field focused on heuristic and rule-based methods. In the 19\textsuperscript{th} century, Launhardt \cite{LAUNHARDT-69,LAUNHARDT-72} introduced one of the earliest mathematical models for cost-minimal layouts, laying the foundation for network-based optimization in civil infrastructure. The 1970s saw the rise of computer-aided approaches, with early models \cite{KIRBY-73,MANDT-73,DYKSTRA-76} integrating terrain data into the road design process. Further advancements in digital elevation models led to tools like PLANS \cite{TWITO-87}, PLANEX \cite{EPSTEIN-99}, and CPLAN \cite{CHUNG-04}, improving road design precision.

Despite these advances, conventional models often imposed simplifying assumptions. For instance, they frequently neglected road curvature, enforced restricted node connectivity, or assumed straight centerlines—limitations that hindered applicability in steep terrains. Stückelberger et al. \cite{STUECKELBERGER-06} highlighted these challenges and demonstrated how classical MST and shortest-path methods were insufficient when roads required curves, switchbacks, or slope feasibility.

Graph theoretic methods have been extensively explored to address these limitations. Algorithms such as Dijkstra’s \cite{DIJKSTRA-59}, A* \cite{PUSHAK-16}, Kruskal’s \cite{KRUSKAL-56}, and Prim’s \cite{PRIM-57} offered robust solutions for shortest paths and minimal spanning trees. The Steiner Minimum Tree (SMT) model, originating from Jakob Steiner’s 19\textsuperscript{th} century geometric investigations into minimal connection networks and later adapted for transportation analysis by Launhardt \cite{LAUNHARDT-72}, provides a foundational framework for optimizing road networks through the strategic inclusion of intermediate junctions that minimize total path length. The computational complexity of the SMT problem was formally established by Karp \cite{KARP-72}, who proved it to be NP-complete, prompting extensive research into both exact and approximate solution approaches.

Subsequent studies introduced Mixed-Integer Linear Programming (MILP) based SMT formulations employing network-flow models \cite{MACULAN-87, GOEMANS-93, KOCH-98}; however, their dependence on flow variables often led to significant computational complexity.
To address this, the present work builds upon flow-free Integer Linear Programming (ILP) formulations that enforce tree topology more efficiently, particularly in the context of terrain-aware infrastructure planning.

Parallel to graph theory, grid-based methods gained popularity. Distance transform (DT) algorithms by de Smith \cite{DESMITH-06} and Li et al. \cite{LI-16} incorporated curvature and slope constraints into bidirectional scanning strategies. Pu et al. \cite{PU-19} extended these ideas to 3D grids, enabling more realistic modeling of rugged terrain. Such approaches are particularly effective for corridor-level route selection, especially when precise geometric feasibility is critical.

HA optimization has been approached from various angles. Some studies leveraged the calculus of variations to derive optimal curves between two points \cite{WAN-95, JHA-06}, but these required simplifying assumptions. Network optimization strategies \cite{TURNER-71, TURNER-78} discretized design spaces into nodes and arcs, but often produced unsmooth alignments. Dynamic programming methods \cite{JHA-06} offered improved feasibility but remained computationally intensive for large-scale problems.

With the rise of computational optimization, more advanced methods emerged. Mondal et al. \cite{MONDAL-15} introduced a bilevel formulation where HA was optimized using black-box solvers like NOMAD \cite{LEDIGABEL-11} and HOPSPACK \cite{GRAY-08}, while VA was solved as a MILP. This structure provided a framework for exact optimization at the lower level but remained dependent on heuristic performance at the upper level.

VA optimization has similarly evolved. Early approaches used LP to improve over graphical mass balance methods \cite{STARK-72}, followed by MILPs that incorporated detailed earthwork cost modeling \cite{MAYER-81, MOREB-96}. Later refinements introduced QNF models with side slopes and haul paths \cite{BEIRANVAND-17, AYMAN-23}, and more recent research shifted toward convex QCQP formulations \cite{MOMO-23, SADHUKHAN-24} to improve scalability.

Three-dimensional alignment optimization (joint HA+VA) remains one of the most challenging problems. Early models based on optimal control theory and splines \cite{CHEW-89} ensured geometric feasibility but were computationally expensive. Subsequent work used genetic algorithms \cite{TAT-03}, simulated annealing \cite{AKAY-06}, and tabu search \cite{ARUGA-05} to address larger terrains. Evolutionary frameworks \cite{JHA-03, JHA-06}, decision-support systems \cite{CHENG-06}, and deep reinforcement learning \cite{GAO-22, HE-23, SONG-23} have also been explored.

Despite this rich body of work, most existing methods either rely on heuristics with no optimality guarantees or struggle to scale efficiently. This motivates the development of deterministic, exact optimization frameworks—like TriPhase—that balance model fidelity, computational tractability, and practical applicability to wind farm road design.

\section{Solution Approach: TriPhase Optimization Model}

To bridge this gap, we present the TriPhase model, a structured three phase optimization pipeline that minimizes wind farm road construction costs while satisfying terrain and engineering constraints. Each phase is tailored to a specific layer of the design hierarchy:

\begin{itemize}
    \item \textbf{Phase 1 – Corridor Selection (CS):} 
    The terrain is discretized into a grid graph where edges are weighted based on slope, curvature, and construction feasibility. We formulate the CS problem as a Steiner minimum tree (SMT) and develop a flow-free ILP model that guarantees tree topology without relying on network flow variables. This improves computational efficiency and scalability while preserving global connectivity.

    \item \textbf{Phase 2 – Segment-wise Horizontal Alignment Optimization (HA):} 
    Each edge from the SMT is treated as a road segment. For each, we apply a bilevel optimization model where the upper level selects discrete horizontal offsets, and the lower level evaluates the corresponding vertical alignment cost using a mixed-integer linear program (MILP). Our method reformulates the previously black-box HA optimization (e.g., Mondal et al.~\cite{MONDAL-15}) into a fully explicit MILP compatible with the Gurobi solver.

    \item \textbf{Phase 3 – Global Vertical Alignment Refinement (VA):} 
    Once horizontal geometry is fixed, we apply a network-wide QCQP model (from Sadhukhan et al.~\cite{SADHUKHAN-24}) to optimize elevation profiles across all road segments. The model balances excavation and embankment volumes while respecting slope constraints, ensuring geometric feasibility and smooth transitions across junctions.
\end{itemize}

We validate the TriPhase model on six wind farm sites using real-world DEM data. Compared to manual designs, our method reduces road construction costs by up to 14\%, while ensuring terrain feasibility and reducing design time through full automation.

\begin{sidewaysfigure}
    \centering
    \includegraphics[scale=0.065]{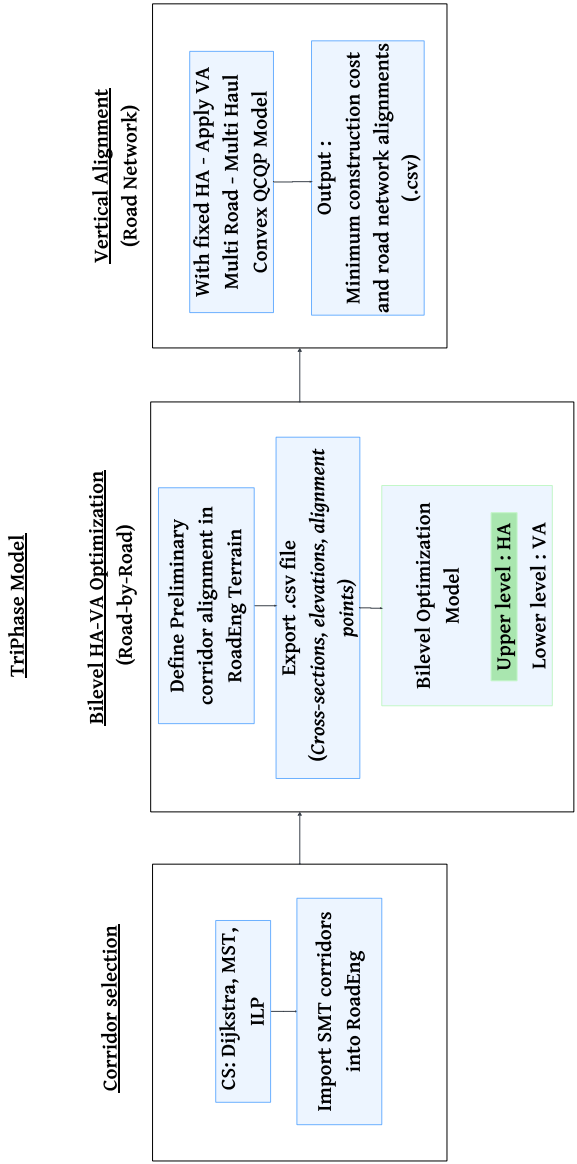}
    \caption{Flowchart of the TriPhase Design workflow.}
    \label{fig:triphase-workflow}
\end{sidewaysfigure}
\subsection{Terrain Data Prepapration}

The terrain data used in this study comes from high-resolution DEMs derived from LiDAR point clouds. We download \texttt{.las} files from OpenTopography \href{https://opentopography.org/}{[OpenTopography]} for each selected wind farm site. DEMs are generated by filtering out non-ground points and extracting earth elevation data required for road design. Since we only have elevation data and no survey, we restrict the experiments to single material. (With appropriate survey information, multimaterial information can be added to our model.)

The raw elevation data is imported into RoadEng\textsuperscript{\textregistered} Terrain software to create detailed surface models. Using RoadEng’s hydrological analysis tools, we identify streams and waterbodies that should be avoided during road alignment. We then overlay turbine locations onto the surface using geospatial shapefiles, preparing the model for subsequent CS and design steps. This prepared terrain model is now ready to be used in RoadEng\textsuperscript{\textregistered} for manually designing the road network. For the TriPhase model, the same data is exported as a \texttt{.csv} file to serve as input for the optimization-based workflow.

\subsection{Phase I: Corridor Selection}

Following terrain preparation, the next critical step is to identify a feasible set of candidate paths from which an optimal road network can be constructed. This process, known as corridor selection (CS), is essential for designing a cost-efficient and geometrically viable road network that connects all turbines to the access point while respecting terrain and hydrological constraints. Rather than manually tracing paths over the surface, we adopt a graph-based optimization framework to systematically explore and evaluate feasible connections. 

The corridor selection phase serves two primary purposes: (1) it reduces the immense search space of possible alignments by restricting design to a filtered subset of terrain-aware road segments, and (2) it provides a mathematically structured input for subsequent optimization steps. In this work, we formulate CS as a Steiner minimum tree (SMT) problem over the preprocessed graph, connecting all turbine locations to the access point through a minimal-cost, terrain-feasible network.

\subsubsection{Discretization of the map}

\begin{figure}[ht]
    \centering
    \includegraphics[scale=0.30]{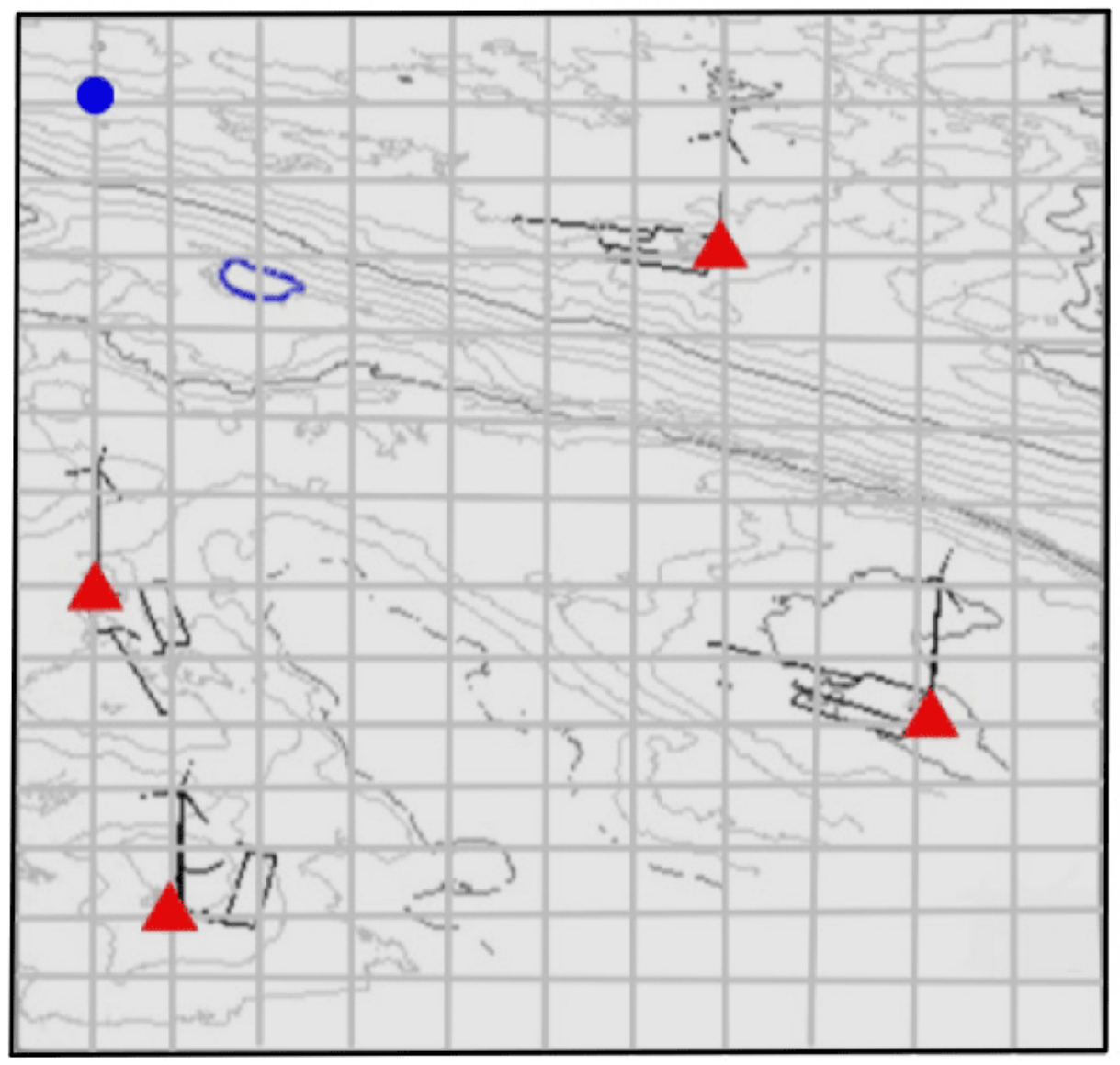}
    \caption{Discretized Map Representation. Intersection of grid lines represent graph nodes, red triangles indicate turbine location, the blue dot indicates the access point. The blue ellipse represents a water body (e.g., a lake), which is considered a forbidden region for road construction.}
    \label{fig:DiscrMap}
\end{figure}

A geographical area is continuous, allowing for an infinite number of possible locations for road connections. However, graph-based optimization methods require a discrete representation of the terrain, where a finite set of nodes and edges define potential road alignments. This transformation from a continuous space to a discrete network is achieved by discretization, in which the terrain is subdivided into a structured grid (Figure \ref{fig:DiscrMap}) of uniform geometric elements.

A resolution of 100m $\times$ 100m is adopted to ensure compatibility with road design constraints, such as the minimum length of curve tangents, straight road segments between curves, and allowable switchback radii. A coarser resolution (e.g., 200m) would reduce the number of nodes but may overlook critical terrain features such as steep slopes, narrow valleys, or localized elevation changes that significantly impact road feasibility and cost. On the other hand, using a finer resolution (e.g., 25m or 50m) would capture more terrain detail but also create a much larger graph with many more nodes and connections. This makes the optimization problem harder to solve and increases the time required for computation, sometimes by several hours, especially for larger wind farm sites.  Our experimentation has found that a 100m $\times$ 100m resolution provides the best balance between computational efficiency and terrain fidelity for the study sites, with cost deviations of less than 4\% compared to the 25 m baseline for the largest windfarm site.

Each node corresponds to a grid line intersection point, providing a structured framework for modeling potential road connections. This discretization approach strikes a balance between computational efficiency and the geometric accuracy required for designing wind farm access roads in complex terrain.

\subsubsection{Nodes and edges in the graph}

In the graph representation of the CS problem, nodes correspond to grid line intersections. Among these, specific nodes are designated as turbine locations and the access point. Edges represent potential road connections between nodes, forming a network that can be optimized for cost and feasibility.

The goal is to construct a connected subgraph that links all turbine locations to the access point while minimizing construction costs. Each edge in the graph is assigned a weight \( W_{(i,j)} \) that reflects the total cost of building that road segment, incorporating factors such as earthwork, pavement, and long-term maintenance. The result is an undirected connected planar graph.

Each edge represents a connection between two adjacent nodes in the grid graph. A road segment is a path composed of one or more edges that connects two key locations, such as between a turbine and the access point.

\subsubsection{Link pattern}

\begin{figure}[ht]
    \centering
    \begin{subfigure}{0.32\textwidth}
        \centering
        \includegraphics[width=\linewidth]{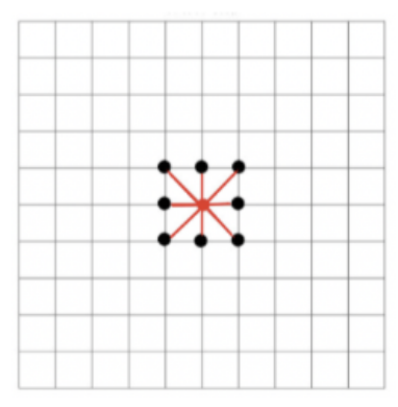}
        \caption{8-link pattern.}
        \label{fig:2a}
    \end{subfigure}
    \hfill
    \begin{subfigure}{0.32\textwidth}
        \centering
        \includegraphics[width=\linewidth]{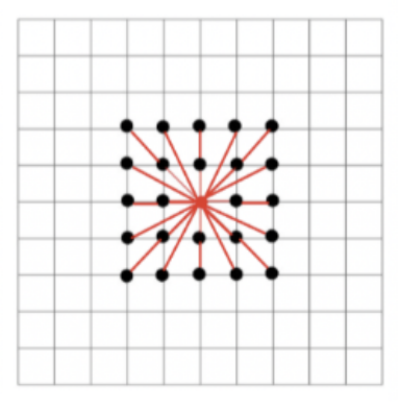}
        \caption{24-link pattern.}
        \label{fig:2b}
    \end{subfigure}
    \hfill
    \begin{subfigure}{0.32\textwidth}
        \centering
        \includegraphics[width=\linewidth]{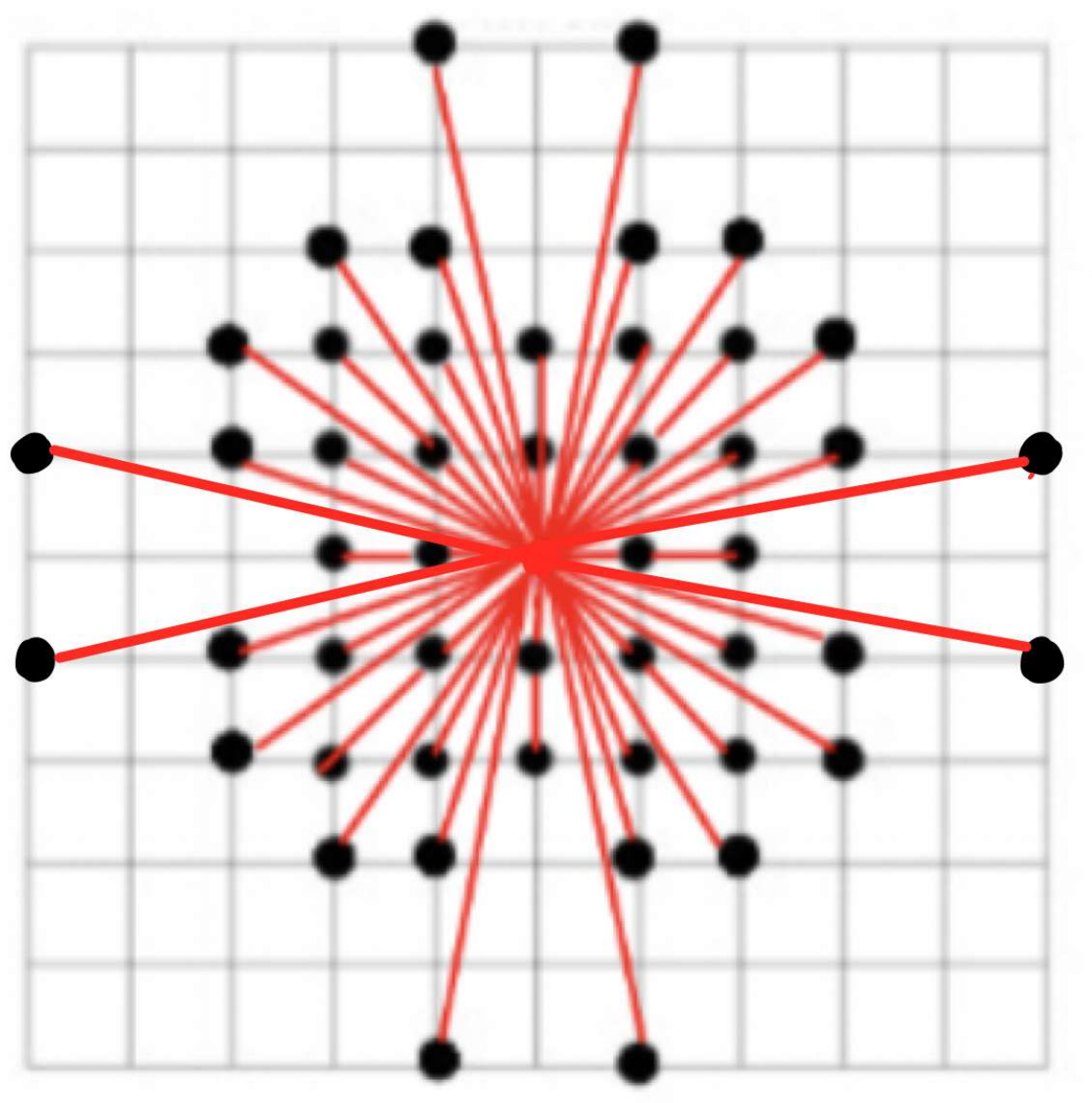}
        \caption{48-link pattern.}
        \label{fig:2c}
    \end{subfigure}

    \caption{Link patterns used at a 100\,m \(\times\) 100\,m grid resolution.}
    \label{fig:linkpatterns}
\end{figure}

The link pattern defines the connectivity structure between nodes in the graph representation of the road network. Each node \( i \in N \) is connected to a predefined set of neighboring nodes, ensuring feasible road alignments while maintaining computational efficiency. The set of allowable connections for each node is denoted as
\begin{align*}
    L(i) = \{ j \in N \mid e_{ij} \in E \},
\end{align*}
where \( e_{ij} \) represents an edge between nodes \( i \) and \( j \). To balance computational efficiency and alignment flexibility, an 8-link connectivity pattern is used. In this model, each node connects to its immediate orthogonal and diagonal neighbors. This structure allows for smooth road transitions while keeping the optimization problem tractable. Although higher degree connectivity models, such as 24-link patterns (Figure \ref{fig:2b}) or 48-link patterns (Figure \ref{fig:2c}) offer increased flexibility~\cite{RETZLAFF-24, NADERIALIZADEH-18, HARDY-23}, they significantly increase computational complexity without substantial improvements in solution quality for this application, as the road network will be further refined at the HA stage. In the experiments that we conducted, 8-link pattern (Figure \ref{fig:2a}) provides an effective compromise by allowing sufficient alignment flexibility while maintaining computational efficiency.

In addition to predefined connectivity, link feasibility is further constrained by gradient and curvature restrictions. Road segments exceeding allowable slope or curvature thresholds are removed before optimization, ensuring that only constructible road links are considered in the SMT formulation. The final connectivity structure is then optimized to minimize overall road construction cost while maintaining network feasibility.

\subsubsection{Gradient constraint}
Gradient constraints ensure that road networks remain feasible for construction and safe for vehicle operations. The gradient of an edge is defined as the ratio of the elevation difference to the horizontal distance. If the gradient exceeds the maximum allowable limit \( g_{max} \), the link is considered infeasible.

For a given edge between nodes \( i \) and \( j \), the gradient is computed as
\begin{equation*}
    g_{ij} = \frac{z_j - z_i}{d_{ij}}
\end{equation*}
where \( z_i, z_j \) are the elevations of nodes \( i \) and \( j \) while \( d_{ij} \) is the horizontal distance between the two nodes.
A road segment is considered feasible if
\begin{equation*}
    |g_{ij}| \leq g_{max}
\end{equation*}
where \( g_{max} \) is the maximum allowable road gradient; it ensures that road alignments remain constructible and safe for vehicle navigation.

In special cases, such as switchbacks where the road alignment changes direction significantly, stricter constraints are applied. When the change in direction \( \Delta \theta \) exceeds a threshold \( \theta_{\text{switchback}} \), the maximum allowable gradient is reduced
\[
g_{link} \leq 0.5 \cdot g_{max}
\]
ensuring that steep paths remain navigable and safe for construction.

\subsubsection{Road network cost function}

Graph-based optimization methods for road network design require a cost function to evaluate the feasibility of each potential link. In this study, the cost of every edge \( W_{(i,j)} \) is determined based on terrain-dependent earthwork, pavement, and maintenance costs. This provides a more accurate representation of construction effort in complex environments.

A gradient-sensitive cost model is employed, where the variability in cost is influenced by local terrain slope. Unlike traditional models that assume uniform costs regardless of location, our formulation accounts for site-specific variations in excavation, embankment, and pavement requirements. Steeper slopes typically lead to increased earthwork due to higher cut and fill volumes and also influence pavement needs and long-term maintenance strategies.

Earthwork costs \( C_{\text{emb}} \) include excavation and embankment operations. Pavement costs \( C_{\text{pav}} \) incorporate the road surface area and widening needed for curves. Maintenance costs \( C_{\text{main}} \) reflect long-term upkeep and are influenced by road gradient and slope exposure. Drainage costs, though significant in forest road planning, are excluded here. Wind farm roads generally follow engineered layouts with stabilized surfaces that do not require extensive drainage \cite{MOSETTI-94, WAN-12}. The cost equations used in this study are adapted from the engineering-based methodology proposed by Ghajar et al.~\cite{GHAJAR-13} who developed a detailed cost estimation framework for forest road construction. Cost components not relevant to wind farm applications are excluded from our formulation. 

\subsubsection*{Earthwork cost}
The cost of excavation and embankment is influenced by the terrain slope $\eta$, which affects both the cut and fill volumes. These volumes are estimated using cross-sectional geometry, ensuring a balance between terrain constraints and earth-moving costs. The excavation volume is estimated as
\begin{align*}
    V_{\text{cut}} &= \frac{w_{\text{cut}}^2 \cdot \tan(\phi_{\text{cut}}) \cdot \tan(\eta)}{2 (\tan(\phi_{\text{cut}}) - \tan(\eta))},\\
    V_{\text{fill}} &= \frac{(w - w_{\text{cut}})^2 \cdot \tan(\phi_{\text{fill}}) \cdot \tan(\eta)}{2 (\tan(\phi_{\text{fill}}) - \tan(\eta))},
\end{align*}
where \( w_{\text{cut}} \) represents the cut width of the edge, \( \phi_{\text{cut}} \) and \( \phi_{\text{fill}} \) are the cut and fill slope angles that vary based on terrain conditions, and \( \eta \) denotes the local slope gradient of the terrain. Using these volumes, the total earthwork cost is computed as
\begin{align}
    C_{\text{emb}} = V_{\text{fill}} \cdot c_{\text{comp}} + V_{\text{cut}} \cdot c_{\text{exc}},\label{eq:emb_cost}
\end{align}
where \( V_{\text{cut}} \) and \( V_{\text{fill}} \) represent the computed cut and fill volumes, respectively. The parameter \( c_{\text{comp}} \) denotes the cost per unit volume for compaction, while \( c_{\text{exc}} \) corresponds to the cost per unit volume for excavation.

\subsubsection*{Pavement cost}  
The pavement cost is influenced by road width, curve radius, and widening requirements at switchbacks. It is given as

\begin{align}
        C_{\text{pav}} = \left( w_{s0} + \frac{k_{\text{cw}}}{r} \right) \cdot l \cdot c_{\text{pav}},\label{eq:pav_cost}
\end{align}
where \( w_{s0} \) represents the standard road width, \( k_{\text{cw}} \) accounts for the additional width required for curves, \( r \) denotes the curve radius, and \( l \) corresponds to the length of the edge. The parameter \( c_{\text{pav}} \) represents the cost per unit area of pavement.

\subsubsection*{Maintenance cost}  
Maintenance costs are derived based on the road gradient $\eta$, assuming that steeper slopes require more frequent and intensive maintenance. The total maintenance cost is computed using

\[
    C_{\text{ann}} = C_{\text{reg}} + \frac{C_{\text{peri}}}{n},
\]
and
\begin{align}
    C_{\text{main}} = C_{\text{con}} + C_{\text{ann}} \cdot \left( \frac{1 - (1 + \text{intr})^{-N}}{\text{intr}} \right),\label{eq:main_cost}
\end{align}
where \( C_{\text{main}} \) is the total maintenance cost over the road’s lifespan, \( C_{\text{con}} \) is the initial construction cost, and \( C_{\text{ann}} \) is the equivalent uniform annual maintenance cost. The annual maintenance cost \( C_{\text{ann}} \) is composed of a regular maintenance component \( C_{\text{reg}} \), and a periodic maintenance component \( C_{\text{peri}} \), which is distributed over a cycle of \( n = 5 \) years. The term intr represents the annual interest rate (with a default value of 2\%), and \( N \) denotes the road amortization period (with a default value of 50 years).

\subsubsection*{Corridor construction cost}

For a given edge \( i \) and \( j \), the total construction cost is defined as:
\begin{align}
    W_{(i,j)} = \alpha \, C_{\text{emb}_{(i,j)}} + \beta \, C_{\text{pav}_{(i,j)}} + \gamma \, C_{\text{main}_{(i,j)}},\label{eq:cost_function}
\end{align}
where \( W_{(i,j)} \) denotes the total cost of constructing the edge \( i \) and \( j \). To see how the earthwork $C_{\text{emb}}$, pavement $C_{\text{pav}}$, and maintenance  cost $C_{\text{main}}$ are calculated, refer to Equations~\eqref{eq:emb_cost}, \eqref{eq:pav_cost}, and \eqref{eq:main_cost}, respectively. The coefficients \( \alpha \), \( \beta \), and \( \gamma \) are user-defined weighting factors that determine the relative importance of each cost component based on project-specific design priorities.

\paragraph{Weight function}

Designing road networks in rugged or constrained environments often requires a balance between geometric feasibility and overall connectivity. Strict enforcement of gradient and curvature constraints can limit the available routing options, especially in areas with steep terrain or limited spatial flexibility. To maintain feasibility while still encouraging geometrically desirable paths, the model incorporates a penalty-based approach. Rather than outright rejecting segments that violate slope or curvature limits, the cost function increases their associated edge weights. This discourages their use in the final network but preserves flexibility by allowing the optimization to consider them when necessary.

\subsection{Phase II: Horizontal Alignment Optimization}

The second phase of the TriPhase model performs horizontal alignment (HA) optimization for each road segment identified in the corridor selection phase. We adapt the bilevel optimization framework introduced by Mondal et al.~\cite{MONDAL-15}, in which the upper level selects a feasible horizontal alignment, while the lower level evaluates its associated vertical alignment cost.

The original model was implemented using black-box derivative-free optimization. In our work, we reformulate the bilevel structure into an explicit mathematical program and implement it using MATLAB and the Gurobi solver.

\subsubsection*{Workflow of Bilevel HA optimization}
\begin{figure}[ht]
    \centering
    \includegraphics[scale=0.07]{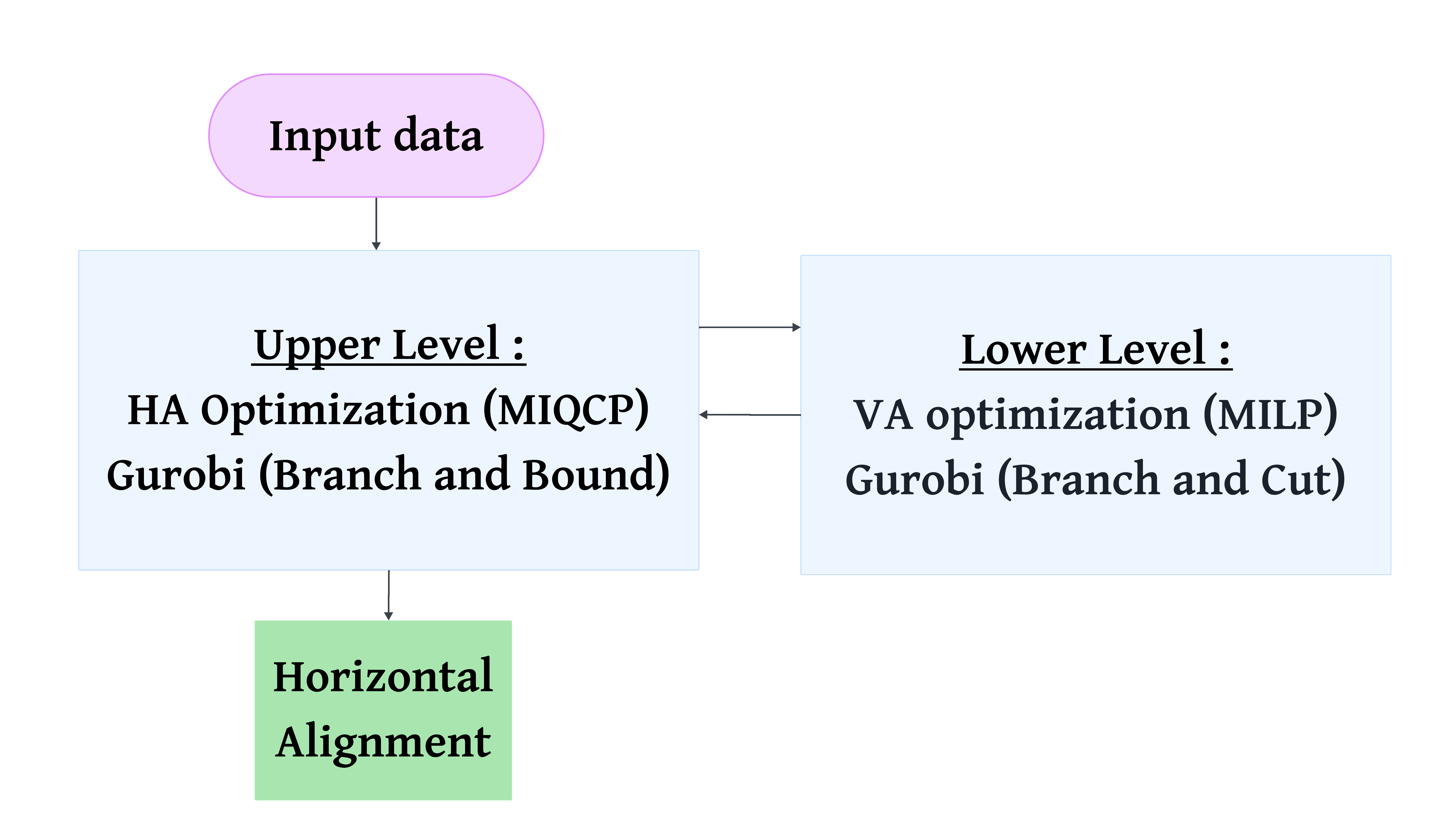}
    \caption{Workflow of the bilevel optimization framework implemented using Gurobi and MATLAB.}
    \label{fig:ha_va_flowchart}
\end{figure}

As illustrated in Figure~\ref{fig:ha_va_flowchart}, the workflow begins by reading and parsing three input files that define the alignment coordinates, cross-sectional terrain areas, and engineering specifications. These inputs collectively describe the geometry, elevation, and construction constraints relevant to the road design problem.

The HA is modeled as a mixed-integer quadratically constrained program, where the decision variables correspond to the coordinates and curvature of the intersection points. The Gurobi solver evaluates candidate alignments by proposing values for these variables. For each candidate, a custom callback function is invoked to simulate the lower level VA model.

Inside the callback, the horizontal geometry is reconstructed using the current decision variables. This involves computing tangent arcs, identifying transition points, and generating cross-sections along the proposed alignment. The vertical profile at each cross-section is then extracted, including material distribution, ground elevation, and design offsets.

These preprocessed vertical profiles form the input to the vertical alignment model, formulated as a MILP. The model determines the optimal elevation level for each station such that earthwork costs are minimized while satisfying constraints on grade and slope transitions.

The resulting cost is returned to the HA model, where it is used as the objective value for the current iteration. Gurobi applies its branch and bound algorithm to explore the feasible space of horizontal alignments, repeatedly invoking the callback until an optimal HA is found. The final output includes the horizontal geometry that minimizes total earthwork cost, along with the corresponding construction cost

A mathematical summary of the HA model is provided in the Appendix \ref{ha-appendix}, and the complete formulation can be found in the original framework \cite{MONDAL-15}.

\subsection{Phase III: Vertical Alignment}

In the final phase of the TriPhase pipeline, the vertical alignment (VA) is optimized across the entire road network, after the horizontal geometry has been fixed. This global optimization ensures smooth transitions at intersections, balanced cut-and-fill volumes, and construction feasibility over complex terrain.

We adopt an existing network-wide model, denoted by $\text{VA}_{\text{network}}$, developed by thesis~\cite{SADHUKHAN-24}. This model formulates vertical alignment optimization as a convex quadratically constrained quadratic program (QCQP), extending the classical quasi-network flow (QNF) formulation. In addition to preforming simultaneous optimization across all roads and intersections in the network, it integrates several realistic engineering elements:

\begin{itemize}
    \item Multiple material types, each with distinct excavation and embankment costs;
    \item Three haul types (short, medium, long), with individualized hauling and loading costs;
    \item Quadratic volume computation based on trapezoidal cross-sectional geometry with side slopes;
    \item Flow-based modeling of material allocation, hauling logistics, and node balance.
\end{itemize}

This formulation is convex and thus guarantees global optimality. It is efficiently solved using Gurobi’s QCQP solver and allows for accurate volume approximation through least-squares-fitted quadratic equations, improving earthwork fidelity without introducing integer variables.

In addition to this final network-wide model, we integrate a simplified MILP-based VA model (see Appendix~\ref{va-single}) into the HA optimization stage. This lightweight formulation estimates segment-wise earthwork costs rapidly and enables efficient evaluation during bilevel HA optimization. Once all segments are assembled into a full network, the comprehensive $\text{VA}_{\text{network}}$ model is applied to compute the final elevation profile.

The $\text{VA}_{\text{network}}$ formulation is fully integrated into our TriPhase optimization framework. For completeness, we include a detailed description of the model in Appendix~\ref{va-network}, with full credit to the original author~\cite{SADHUKHAN-24}.

\section{Results}\label{sec5}

Numerical experiments are conducted to evaluate the performance of the proposed \emph{TriPhase} road optimization framework, which integrates CS, HA, and VA. The framework is evaluated on six real-world wind farm sites; five publicly available projects and one additional site provided by Softree Technical Systems Inc. For each site, total construction costs are compared against designs generated using Softree’s \textit{RoadEng} software (version 11.0). In addition, the resulting road networks are visualized and computational times for each optimization phase are reported.

The six test sites span a wide range of project scales and terrain complexities. Key site characteristics, including the number of turbines, HA intersection points, VA stations, and total site area, are summarized in Table~\ref{table:site-overview}. These cases are used to assess the robustness and scalability of the proposed three-stage optimization framework.

\begin{table}[tbph]
\centering
\setlength{\tabcolsep}{4pt} 
\caption{Overview of test sites used for numerical experiments, listing number of turbines, HA intersection points, VA stations, and site area.\label{table:site-overview}}
\begin{tabular}{@{}lrrrr@{}} \toprule
\textbf{Site} & \textbf{\# Turbines} & \textbf{\# IPs} & \textbf{\# Stations} & \textbf{Area} \\
             &                      & (HA)            & (VA)                 & (km\textsuperscript{2}) \\ \midrule
Toksook Bay          & 3  & 5   & 17    & 2.1 \\
Hopkins Ridge II     & 4  & 9   & 42    & 3.0 \\
Prospector Project   & 5  & 23  & 67    & 3.9 \\
Site J               & 7  & 32  & 188   & 11.0 \\
Golden Valley        & 8  & 14  & 131   & 4.5 \\
Midland Wind Project & 25 & 98  & 1,246 & 168.0 \\ \bottomrule
\end{tabular}
\end{table}

\subsection{Experimental Setup}
All numerical experiments were implemented in MATLAB R2024b \cite{MATLAB-24b} using YALMIP \cite{LOFBERG-04} for model formulation and Gurobi Optimizer \cite{GUROBI-25} for mixed-integer and quadratic optimization. The CS phase was implemented from scratch and solved as an ILP-based SMT optimization. The HA phase follows an explicit segment-wise formulation, with each road segment optimized independently using branch-and-bound. The VA phase employs a convex QCQP formulation applied over the complete road network.

All computations were performed on a Linux-based workstation running Ubuntu~25.04 \cite{UBUNTU-25}, equipped with dual Intel Xeon Gold 6426Y processors, NVIDIA RTX A6000 GPUs, and 256~GB of RAM. GPU acceleration was utilized during the solution of the CS model, enabling efficient solution of large-scale instances and consistent solver performance across all test sites.

\subsection{Construction cost comparison}

Across all test sites, the TriPhase framework consistently yields lower estimated construction costs than the manual design approach (refer Table \ref{table:cost-comparison}). Both sets of alignments are evaluated using the same RoadEng cost engine and identical material assumptions, the observed differences reflect improvements in network geometry.

Two distinct trends emerge from the results. For sites with relatively simple terrain and limited geometric constraints, the cost difference between the two approaches is minimal, with reductions below 1\%. In contrast, for larger and more complex sites, the TriPhase framework achieves substantially lower construction costs, with reductions exceeding 10\% in some cases. For example, at the Midland Wind Project, the optimized design reduces the total estimated cost by approximately \$0.6~million compared to the manual layout.

These differences can be attributed to the integrated nature of the TriPhase framework, which jointly optimizes CS, HA, and VA across the entire road network. By extending the design process beyond sequential manual workflows, the framework enables more effective global trade-offs, resulting in greater cost savings as site complexity increases.

\begin{table}[tbph]
\centering
\caption{Comparison of construction costs across test sites.}
\label{table:cost-comparison}
\resizebox{\columnwidth}{!}{%
\begin{tabular}{lcrrr}
\toprule
\textbf{Site} & \textbf{\# Turbines} & \multicolumn{2}{c}{\textbf{Construction Cost (\$1{,}000s)}} & \textbf{Diff (\%)} \\
\cmidrule(lr){3-4}
 & & \textbf{Manual Design} & \textbf{TriPhase Model} & \\
\midrule
TB     & 3  & 345.6   & 343.3   & 0.7  \\
HR II  & 4  & 433.4   & 433.2   & 0.0  \\
PP     & 5  & 957.8   & 885.0   & 7.6  \\
Site J & 7  & 1{,}217.9 & 1{,}109.1 & 8.9  \\
GV     & 8  & 1{,}004.2 & 956.3   & 4.8  \\
MWP    & 25 & 4{,}561.2 & 3{,}928.2 & 13.9 \\
\bottomrule
\end{tabular}}
\end{table}

\subsection{Visualization and road network comparison}

The manual design workflow combines corridor selection and horizontal alignment into a single process, hence it is not possible to visualize the corridor selection stage independently. The visual comparison therefore focuses on the final road networks produced by the two approaches, which implicitly capture differences in how corridors are selected and refined.

\begin{figure}[tb]
\centering
\includegraphics[width=\columnwidth]{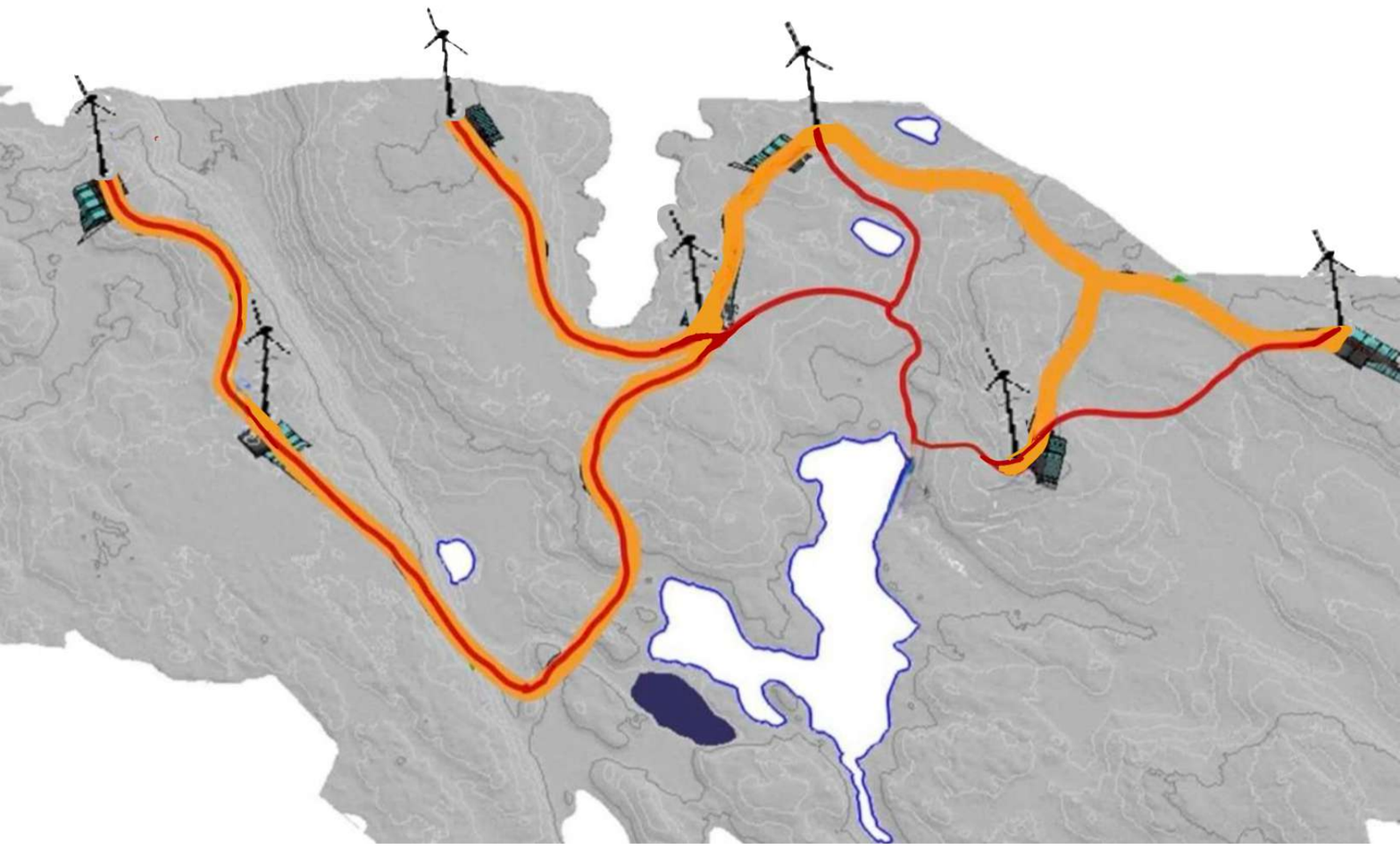}
\caption{Site J: Alignment comparison for a 7-turbine site. Black-outlined icons denote turbines. The yellow alignment represents the manual design, while the red alignment corresponds to the TriPhase result.}
\label{fig:SiteJ}
\end{figure}

Figure~\ref{fig:SiteJ} presents a representative comparison for Site~J, a seven-turbine site characterized by complex terrain and multiple routing alternatives. The manual design and TriPhase solutions exhibit noticeable differences in overall network geometry, reflecting distinct routing decisions made during the design process. In particular, the TriPhase solution explores alternative corridors that are not selected in the manual workflow, resulting in a road network with smoother transitions and more efficient curvature.

These geometric differences translate into reduced construction costs while maintaining feasibility across challenging terrain. Although the optimized alignment differs substantially from the manual layout in certain segments, it remains practical and constructible, demonstrating the ability of the TriPhase framework to identify cost-effective routing strategies that may be difficult to obtain through manual design alone. This example illustrates the practical value of optimization-driven road network design for complex wind farm sites.

\subsection{Key takeaways: Impact of corridor selection}

The experimental results yield the following key insights into the influence of corridor selection on solution quality and computational performance:
\begin{itemize}
    \item Although the Steiner-based corridor selection problem is NP-hard in theory, it is practically tractable for real-world wind farm instances. Small and medium-sized sites are solved within minutes, while even large-scale sites remain solvable within operationally acceptable time limits.
    \item The computational cost of corridor selection increases rapidly with problem size and becomes the primary runtime bottleneck for large sites; however, this growth can be substantially mitigated through graph reduction techniques and careful control of spatial resolution.
    \item Using coarser digital elevation model resolutions leads to significant reductions in computation time with only marginal increases in construction cost, demonstrating a favorable trade-off between terrain fidelity and computational efficiency for most sites.
    \item The expanded-graph corridor selection formulation preserves optimal construction costs while achieving up to an order-of-magnitude reduction in solution time, thereby enabling scalable optimization without compromising solution quality.
    \item Corridor selection, in combination with horizontal alignment, accounts for the majority of the observed cost savings, whereas subsequent optimization stages contribute primarily incremental improvements.
\end{itemize}

\section{Conclusions}\label{sec6}

TriPhase introduces a comprehensive three-phase optimization framework for minimizing road construction costs in wind farm development. The model integrates terrain-aware corridor selection using an ILP-based Steiner Minimum Tree, segment-wise horizontal alignment optimization via a bilevel formulation, and global vertical alignment refinement through a convex QCQP model. Together, these stages offer a structured pipeline that captures both the topological and geometric complexity of wind farm road design.

Unlike traditional manual approaches, which are often time-consuming and suboptimal, this framework automates the design process while explicitly enforcing constraints on slope, curvature, and material balance. It operates on real terrain data, incorporates practical construction parameters, and is fully compatible with commercial tools such as Softree RoadEng\textsuperscript{\textregistered}.

Numerical experiments on six wind farm sites demonstrate that TriPhase consistently outperforms manual baselines, achieving up to 14\% reduction in construction cost. The modular design also supports scalability and transparency, with each phase solvable by standard optimization software like Gurobi.

Several extensions remain open for future research, including joint HA optimization across the full network, finer DEM resolutions, and multi-objective formulations that account for emissions, maintenance, or environmental impact. Further integration of geotechnical and hydrological data may also enhance route stability and resilience in steep or erosion-prone terrain.

Overall, TriPhase establishes a practical and extensible foundation for data-driven road design in the renewable energy sector, where infrastructure cost, terrain feasibility, and sustainability are equally critical.

\subsection*{Data availability}
Data used in the numerical experiments is available on demand by contacting the authors and subject to Softree Technical Systems Inc. approval.

\subsection*{AI tools and technology}

Generative AI tools were used only for editing the paper and translation of German sources; all technical content and model development were solely conducted and verified by the authors.

\section*{Acknowledgments}
This research was funded by Innovate BC under Ignite grants [IGNITE-2021-RND11-277-UBCO-Lucet-Softree], and by the Natural Sciences and Engineering Research Council of Canada under Discovery Grant [RGPIN-2018-03928]. The Ignite grant was supported by Softree Technical Systems Inc. (\url{http://www.softree.com}, which was instrumental in providing data (ground maps), a software license for RoadEng, and technical expertise. Data obtained from six wind farm sites was supplied: one by Softree and five from the U.S. Wind Turbine Database.

\appendix

\section{Horizontal Alignment Optimization Model}\label{ha-appendix}

This appendix outlines the MIQCP formulation used to optimize the horizontal alignment (HA) of a single road segment between fixed endpoints. The alignment is defined by horizontal points of intersection (HPIs), denoted by \( P_i = (x_i, y_i) \), for \( i = 0, \dots, n \). The endpoints \( P_0 \) and \( P_n \) are fixed, while the intermediate points and their associated radii \( r_i \) are decision variables constrained by curvature and corridor geometry.

The model is solved using Gurobi’s branch-and-bound algorithm. For each candidate HA configuration, a callback is triggered to evaluate earthwork costs via the vertical alignment MILP model (Appendix~\ref{va-single}). Further details of the original formulation are provided in~\cite{MONDAL-15}.

The objective function minimizes the earthwork cost from the corresponding vertical alignment:
\begin{equation}
\min f(X) = \mathcal{C}_{VA}(X),
\end{equation}
where \( \mathcal{C}_{VA}(X) \) is the total cost returned from solving the vertical alignment MILP for geometry \( X \).

Each interior HPI must lie within a rectangular corridor:
\begin{align}
\underline{b}_{x_i} &\leq x_i \leq \overline{b}_{x_i}, \\
\underline{b}_{y_i} &\leq y_i \leq \overline{b}_{y_i}, \quad \forall i = 1, \dots, n-1.
\end{align}

Curvature feasibility is enforced as:
\begin{equation}
r_i \geq R_{\min}, \quad \forall i = 1, \dots, n-1.
\end{equation}

Convex quadratic constraints ensure geometric continuity:
\begin{align}
\| P_{i} - P_{i-1} \|_2^2 &\geq \| P_{i-1} - F_{i-1} \|_2^2, \\
\| P_{i} - P_{i-1} \|_2^2 &\geq \| P_{i} - E_{i} \|_2^2, \\
\| P_{i+1} - P_i \|_2^2 &\geq \| P_i - F_i \|_2^2, \\
\| P_{i+1} - P_i \|_2^2 &\geq \| P_{i+1} - E_{i+1} \|_2^2, \quad \forall i = 1, \dots, n-2.
\end{align}

Here, \( E_i \) and \( F_i \) represent the entry and exit points of the circular arc at each HPI \( P_i \), derived from the tangent direction and assigned radius \( r_i \). These are precomputed prior to optimization to preserve model convexity and solver compatibility.

\section{Vertical Alignment Model}
\label{VA-models}

\subsection{Mixed Integer Linear Program - Quasi Network Flow Model 
\label{va-single}}

This appendix provides an outline of the quasi-network flow (QNF) MILP model used for solving the VA optimization problem. The detailed explanation of the model can be found in \cite{HARE-14}, and the following summary is intended for clarity and confirmation that the model is indeed a MILP. The variables involved are $f_{ij,t}^k$, $V_i^+$, $V_i^-$, $u_i$, $a_{g,i}$, and binary variables $\nu_{i,l}$, $y_{k,t}$. The objective function is given as
\begin{multline}
\min \sum_{i \in S \cup B} p_i V_i^+ + \sum_{i \in S \cup W} q_i V_i^- \\
+ \sum_{i \in S} \sum_{t \in T} \left( c_{i,t}^r f_{i,i-1,t}^r + c_{i,t}^r f_{i,i+1,t}^r \right) \\
+  \sum_{j \in B} \sum_{t \in T} c_{dj} \left( f_{j, g(j)-1, t}^b + f_{j, g(j)+1, t}^b \right) \\
+ \sum_{j \in W} \sum_{t \in T} c_{dj} \left( f_{\phi(j)-1, j, t}^w + f_{\phi(j)+1, j, t}^w \right).
\end{multline}

The model also includes several linear constraints such as material conservation for all $t \in T, i \in S$,
\begin{equation}
f_{i-1, i, t}^r + f_{i, i-1, t}^r + \sum_{j \in B} f_{j, i, t}^b = f_{i, i+1, t}^r + f_{i+1, i, t}^r + \sum_{j \in W} f_{i, j, t}^w,
\end{equation}
\begin{equation}
f_{i+1, i, t}^r + f_{i, i+1, t}^r + \sum_{j \in B} f_{i, j, t}^b = f_{i, i-1, t}^r + f_{i-1, i, t}^r + \sum_{j \in W} f_{j, i, t}^w,
\end{equation}
and volume balance constraints,
\begin{equation}
\sum_{t \in T} f_{i, i-1, t}^u + f_{i, i+1, t}^u + f_{i, i, t}^u = V_i^+, \quad \forall i \in S,
\end{equation}
\begin{equation}
\sum_{t \in T} f_{i-1, i, t}^l + f_{i+1, i, t}^l + f_{i, i, t}^l = V_i^-, \quad \forall i \in S,
\end{equation}
\begin{equation}
\sum_{t \in T} f_{j, g(j)-1, t}^b + f_{j, g(j)+1, t}^b = V_j^+, \quad \forall j \in B,
\end{equation}
\begin{equation}
\sum_{t \in T} f_{\phi(j)-1, j, t}^w + f_{\phi(j)+1, j, t}^w = V_j^-, \quad \forall j \in W.
\end{equation}

The volume constraints relate to the quadratic spline as
\begin{equation}
P(s) =
\begin{cases}
P_1(s), & \text{if } S \delta(1, 1) \leq s \leq S \delta(1, n_1), \\
P_2(s), & \text{if } S \delta(2, 1) \leq s \leq S \delta(2, n_2), \\
\vdots \\
P_g(s), & \text{if } S \delta(g, 1) \leq s \leq S \delta(g, n_g).
\end{cases}
\end{equation}

To ensure continuity and grade feasibility between adjacent segments, the following constraints are enforced,
\begin{align}
P_{g-1}(S_{g,1}) &= P_g(S_{g,1}), \quad P'_{g-1}(S_{g,1}) = P'_g(S_{g,1}), \notag \\
G_L &\leq P'_g(S_{g,1}) \leq G_U, \quad \forall g \in \mathcal{G} \setminus \{1\}.
\end{align}

\subsection{Quadratically Constrained Quadratic Program - Quasi Network Flow Model}
\label{va-network}

This appendix provides an outline of the \( \text{VA}_{\text{network}} \) model formulated as a convex quadratically constrained quadratic program (QCQP) for solving the VA optimization problem. The detailed model structure is based on the extensions introduced in \cite{SADHUKHAN-24}, and the following summary confirms that the formulation is convex and solver-compatible as a QCQP.

The objective of the MRMH-QNF model is to minimize the total construction cost, which includes excavation, embankment, hauling, and loading. The objective function is defined as follows:

\begin{multline}
\min \Big[ 
\sum_{\substack{i \in \mathcal{R} \\ j \in \mathcal{S}_i \cup \mathcal{B}_i \\ m \in \mathcal{K} \\ h \in \mathcal{H}}} 
(p_{m,h} + y_{m,h}) Y^+_{i,j,m,h} 
+ \sum_{\substack{i \in \mathcal{R} \\ j \in \mathcal{S}_i \cup \mathcal{W}_i \\ m \in \mathcal{K} \\ h \in \mathcal{H}}} 
q_e Y^-_{i,j,m,h} \\
+ \sum_{\substack{i \in \mathcal{R} \\ j \in \mathcal{S}_i \\ m \in \mathcal{K} \\ h \in \mathcal{H}}} 
\left( c^{+,h}_{i,j,m} t^{+,h}_{i,j,m} + c^{-,h}_{i,j,m} t^{-,h}_{i,j,m} \right) 
+ \sum_{\substack{e \in \mathcal{I} \\ m \in \mathcal{K} \\ h \in \mathcal{H}}} 
(p_{m,h} + y_{m,h}) U^+_{e,m,h} \\
+ \sum_{\substack{e \in \mathcal{I} \\ m \in \mathcal{K} \\ h \in \mathcal{H}}} 
q_{m,h} U^-_{e,m,h} 
+ \sum_{\substack{i \in \mathcal{R} \\ i \in \xi(e) \\ j \in \gamma_1(i,e) \\ m \in \mathcal{K} \\ h \in \mathcal{H}}} 
c^{e,h}_{i,j,m} \left( f^{in,e,h}_{i,j,m} + f^{out,e,h}_{i,j,m} \right) \\
+ \sum_{\substack{i \in \mathcal{R} \\ k \in \mathcal{B}_i \\ m \in \mathcal{K} \\ h \in \mathcal{H}}} 
\left( p_{m,h} + y_{m,h} + c_{m,h} \tilde{d}_k \right) 
\left( f^{b+,k,h}_{i,\vartheta(k),m} + f^{b-,k,h}_{i,\vartheta(k),m} \right) \\
+ \sum_{\substack{i \in \mathcal{R} \\ l \in \mathcal{W}_i \\ m \in \mathcal{K} \\ h \in \mathcal{H}}} 
\left( q_{m,h} + c_{m,h} \tilde{d}_l \right) 
\left( f^{w+,l,h}_{i,\varphi(l),m} + f^{w-,l,h}_{i,\varphi(l),m} \right) 
\Big].
\end{multline}

We now present the key constraints that govern VA continuity, elevation offsets, and material flow in the network. These constraints ensure smooth transitions between road segments, consistent elevations at intersections, and physically feasible cut-and-fill operations across all road and junction elements.

\paragraph{Continuity constraints}
To maintain smooth transitions, the elevation and gradient of the initial section of each segment must align with those of the final section of the preceding segment. This continuity is enforced by for all $g \in \mathcal{G}_i \setminus \{1\}, i \in \mathcal{R}$,
\begin{align}
P_{i,g-1}\left(x_{\varphi(i,g-1,n_{i,g-1})}\right) 
&= P_{i,g}\left(x_{\varphi(i,g,1)}\right), \notag \\
P'_{i,g-1}\left(x_{\varphi(i,g-1,n_{i,g-1})}\right) 
&= P'_{i,g}\left(x_{\varphi(i,g,1)}\right).
\end{align}

Likewise, for all road segments converging at an intersection, their elevation values must be consistent at the point of convergence. This is enforced by,
\begin{multline}
P_{\mu(e,1),\eta(e,1)}\Big( 
\ x_{\varphi(\mu(e,1),\eta(e,1),\zeta(e,1))} \Big) \notag \\
= \ 
P_{\mu(e,n_{r_e}),\eta(e,n_{r_e})}\Big( 
\ x_{\varphi(\mu(e,n_{r_e}),\eta(e,n_{r_e}),\zeta(e,n_{r_e}))} \Big), \notag \\
\quad \forall e \in \mathcal{I},\ \forall i \in \mathcal{R}.
\end{multline}
Here, both \( P \) and \( P' \) are functions of \( x \), where the coefficients are treated as optimization variables. These constraints collectively ensure that the resulting vertical profile is both continuous and constructible throughout the network.

\paragraph{Gap Constraints}
To ensure geometric consistency, the model enforces that the gap between the designed road profile and the existing terrain elevation matches the vertical offset variable at each section. For every segment \( g \) and section \( k \) of road \( i \), this condition is enforced by for all $i \in \mathcal{R}$, $j \in \mathcal{S}_i$, $g \in \mathcal{G}_i$, and $k \in \mathcal{S}_{i,g}$,
\begin{equation}
P_{i,g} \left( x_{\varphi(i,g,k)} \right) - Z_{i,j} = u_{i,j}.
\end{equation}
Here, \( Z_{i,j} \) denotes the ground elevation at the \( j^{\text{th}} \) section of road \( i \), and \( u_{i,j} \) represents the corresponding design offset.

At intersections, a consistent vertical offset must be maintained between the intersection node and the connected road segments. This requirement is imposed by,
\begin{equation}
u_e = u_{\xi(e)}, \quad \forall e \in \mathcal{I},
\end{equation}
ensuring that the vertical profile at an intersection agrees with that of the adjacent incoming or outgoing road sections.

\paragraph{Flow Constraints}
To ensure material conservation at virtual transit nodes, the total incoming flow into a section must equal the total outgoing flow. When a section is not connected to an intersection, we impose the following flow balance. For the forward direction, we require for all $i, j, m, h$
\begin{equation}
f_{i,j,m}^{t+,h} + f_{i,j,m}^{l+,h} + \sum_{l \in \mathcal{W}_i} f_{i,\delta(l),m}^{w+,l,h}
= f_{i,j-1,m}^{t+,h} + f_{i,j,m}^{u+,h}
+ \sum_{k \in \mathcal{B}_i} f_{i,\nu(k),m}^{b+,k,h}.
\end{equation}
The backward flow balance is given by for all $i, j, m, h$,
\begin{equation}
f_{i,j,m}^{t-,h} + f_{i,j,m}^{l-,h} + \sum_{l \in \mathcal{W}_i} f_{i,\delta(l),m}^{w-,l,h}
= f_{i,j-1,m}^{t-,h} + f_{i,j,m}^{u-,h}
+ \sum_{k \in \mathcal{B}_i} f_{i,\nu(k),m}^{b-,k,h}.
\end{equation}

In cases where a section is linked to an intersection and is the first segment of a road, the flow constraints become slightly different. For forward flow, we impose for all $i, j, m, h$,
\begin{equation}
f_{i,j,m}^{t+,h} + \sum_{l \in \mathcal{W}_i} f_{i,\delta(l),m}^{w+,l,h}
= f_{i,j,h}^{out,\lambda(i,j)} + \sum_{k \in \mathcal{B}_i} f_{i,\nu(k),m}^{b+,k,h}.
\end{equation}
For the backward direction, the constraint is for all $i, j, m, h$,
\begin{equation}
f_{i,j,m}^{in,\lambda(i,j),h} + \sum_{l \in \mathcal{W}_i} f_{i,\delta(l),m}^{w-,l,h}
= f_{i,j-1,m}^{t-,h} + \sum_{k \in \mathcal{B}_i} f_{i,\nu(k),m}^{b-,k,h}.
\end{equation}

If the section is the last segment of a road and is also connected to an intersection, the flow constraints are adjusted accordingly. For incoming flow at the junction, we require,
\begin{equation}
f_{i,j,m}^{in,\lambda(i,j),h} + \sum_{l \in \mathcal{W}_i} f_{i,\delta(l),m}^{w+,l,h}
= f_{i,j-1,m}^{t+,h} + \sum_{k \in \mathcal{B}_i} f_{i,\nu(k),m}^{b+,k,h}, 
\quad \forall i, j, m, h.
\end{equation}
For outgoing flow,
\begin{equation}
f_{i,j,m}^{t-,h} + \sum_{l \in \mathcal{W}_i} f_{i,\delta(l),m}^{w-,l,h}
= f_{i,j,m}^{out,\lambda(i,j),h} + \sum_{k \in \mathcal{B}_i} f_{i,\nu(k),m}^{b-,k,h}, 
\quad \forall i, j, m, h.
\end{equation}

At each intersection node, we enforce a global conservation constraint to ensure that the total material entering the intersection matches the total material exiting. This is imposed as for all $e \in \mathcal{I}$, $m \in \mathcal{K}$, $h \in \mathcal{H}$,
\begin{equation}
\sum_{i \in \xi(e)} f_{i,T(e,i),m}^{in,e,h} + f_{e,m}^{u+,h}
= \sum_{i \in \xi(e)} f_{i,T(e,i),m}^{out,e,h} + f_{e,m}^{l+,h}.
\end{equation}

Finally, to avoid double-counting, we restrict the cut and fill quantities at sections connected to an intersection, since those materials are already accounted for in the intersection flows. Therefore, we impose for all $e \in \mathcal{I}$, $m \in \mathcal{K}$, $h \in \mathcal{H}$,
\begin{equation}
\sum_{i \in \xi(e)} f_{i,T(e,i),m}^{u+,h} + \sum_{i \in \xi(e)} f_{i,T(e,i),m}^{u-,h} = 0,
\end{equation}
\begin{equation}
\sum_{i \in \xi(e)} f_{i,T(e,i),m}^{l+,h} + \sum_{i \in \xi(e)} f_{i,T(e,i),m}^{l-,h} = 0.
\end{equation}

\paragraph{Balance constraints}
To ensure that material quantities remain consistent, the total unloading flows from a section must match the cut volume, and the total loading flows into the section must match the fill volume. Accordingly, we impose the following balance constraints. For each road section, the unloading balance is given by for all $i \in \mathcal{R}$, $j \in \mathcal{S}_i$, $m \in \mathcal{K}$,
\begin{equation}
\sum_{h \in \mathcal{H}} f_{i,j,m}^{u+} + \sum_{h \in \mathcal{H}} f_{i,j,m}^{u-} 
= \sum_{h \in \mathcal{H}} \mathcal{V}_{i,j,m}^{+}.
\end{equation}
Similarly, the loading balance for the same section is,
\begin{equation}
\sum_{h \in \mathcal{H}} f_{i,j,m}^{l+} + \sum_{h \in \mathcal{H}} f_{i,j,m}^{l-} 
= \sum_{h \in \mathcal{H}} \mathcal{V}_{i,j,m}^{-}.
\end{equation}

A similar consistency requirement applies at each intersection. The unloading volume at an intersection must equal the total excavated quantity stored there, which gives,
\begin{equation}
\sum_{h \in \mathcal{H}} f_{e,m}^{u} = \sum_{h \in \mathcal{H}} \mathcal{U}_{e,m}^{+}, 
\quad \forall e \in \mathcal{I},\ \forall m \in \mathcal{K}.
\end{equation}
For loading at the same intersection, we impose,
\begin{equation}
\sum_{h \in \mathcal{H}} f_{e,m}^{l} = \sum_{h \in \mathcal{H}} \mathcal{U}_{e,m}^{-}, 
\quad \forall e \in \mathcal{I},\ \forall m \in \mathcal{K}.
\end{equation}

\paragraph{Bound constraints}
The model imposes bounds to restrict the domain of all decision variables. Let \( M_{i,j,m}^{+} \) and \( M_{i,j,m}^{-} \) represent the maximum volumes of material \( m \) that can be excavated from and filled into section \( j \) of road \( i \), respectively. Similarly, let \( \bar{u}_{i,j} \) and \( \underline{u}_{i,j} \) denote the upper and lower elevation offsets for section \( j \) of road \( i \), while \( \bar{z}_e \) and \( \underline{z}_e \) represent the maximum and minimum elevation bounds at intersection \( e \). With these definitions, we impose the following constraints.

For excavation and fill flow bounds within each road section, we require for all $i \in \mathcal{R}$, $j \in \mathcal{S}_i$, $m \in \mathcal{K}$, $h \in \mathcal{H}$,
\begin{equation}
0 \leq f_{i,j,m}^{u+,h} \leq M_{i,j,m}^{+,h},
\end{equation}
\begin{equation}
0 \leq f_{i,j,m}^{u-,h} \leq M_{i,j,m}^{+,h},
\end{equation}
\begin{equation}
0 \leq f_{i,j,m}^{l+,h} \leq M_{i,j,m}^{+,h},
\end{equation}
\begin{equation}
0 \leq f_{i,j,m}^{l-,h} \leq M_{i,j,m}^{+,h}.
\end{equation}

For intersection-related material flows, we enforce,
\begin{equation}
0 \leq f_{e,m}^{u+,h} \leq M_{e,m}^{+,h}, 
\quad \forall e \in \mathcal{I},\ \forall m \in \mathcal{K},\ \forall h \in \mathcal{H}.
\end{equation}
\begin{equation}
0 \leq f_{e,m}^{u-,h} \leq M_{e,m}^{-,h}, 
\quad \forall e, m, h.
\end{equation}

The elevation offset variables are bounded as follows,
\begin{equation}
\underline{u}_{i,j} \leq u_{i,j} \leq \bar{u}_{i,j}, 
\quad \forall i \in \mathcal{R},\ \forall j \in \mathcal{S}_i,
\end{equation}
\begin{equation}
\underline{z}_e \leq z_e \leq \bar{z}_e, 
\quad \forall e \in \mathcal{I}.
\end{equation}

To ensure physical feasibility of the volume variables, we also impose,
\begin{equation}
M_{i,j,m}^{+} \geq \mathcal{V}_{i,j,m}^{+,h} \geq 0,\quad 
M_{i,j,m}^{-} \geq \mathcal{V}_{i,j,m}^{-,h} \geq 0, 
\quad \forall i, j, m, h.
\end{equation}
\begin{equation}
M_{e,m}^{+} \geq \mathcal{U}_{e,m}^{+,h} \geq 0,\quad 
M_{e,m}^{-} \geq \mathcal{U}_{e,m}^{-,h} \geq 0, 
\quad \forall e, m, h.
\end{equation}

For all flow variables related to borrow and waste points, we include,
\begin{equation}
f_{i,\nu(k),m}^{b+,k,h} \geq 0, \quad 
f_{i,\nu(k),m}^{b-,k,h} \geq 0, 
\quad \forall i \in \mathcal{R},\ \forall k \in \mathcal{B}_i,\ \forall m,\ h.
\end{equation}
\begin{equation}
f_{i,\delta(l),m}^{w+,l,h} \geq 0, \quad 
f_{i,\delta(l),m}^{w-,l,h} \geq 0, 
\quad \forall i \in \mathcal{R},\ \forall l \in \mathcal{W}_i,\ \forall m,\ h.
\end{equation}

Finally, for internal material transit variables across sections and intersections, we require,
\begin{equation}
f_{i,j,m}^{t+,h} \geq 0, \quad 
f_{i,j,m}^{t-,h} \geq 0, 
\quad \forall i \in \mathcal{R},\ \forall j \in \mathcal{S}_i,\ \forall m,\ h.
\end{equation}
\begin{equation}
f_{i,j,m}^{in,e,h} \geq 0, \quad 
f_{i,j,m}^{out,e,h} \geq 0, 
\quad \forall i \in \mathcal{R},\ \forall j \in \mathcal{S}_i,\ \forall e \in \mathcal{I},\ \forall m,\ h.
\end{equation}

\paragraph{Quadratic Least Squares Constraints}
To incorporate volume constraints for multiple materials, we begin with the linear least squares approximation for cut and fill volumes. When a section is in cut (\( u_{i,j,m} \leq 0 \)), the estimated volume must satisfy,
\begin{equation}
\mathcal{V}_{i,j,m}^{+} \geq \chi_{i,j,m,1} u_{i,j,m} + \chi_{i,j,m,2}, 
\quad \forall i \in \mathcal{R},\ \forall j \in \mathcal{S}_i,\ \forall m \in \mathcal{K}.
\end{equation}
When in fill (\( 0 \leq u_{i,j,m} \leq \bar{u}_{i,j,m} \)), the constraint becomes,
\begin{equation}
\mathcal{V}_{i,j,m}^{-} \geq \chi_{i,j,m,1} u_{i,j,m} + \chi_{i,j,m,2}, 
\quad \forall i \in \mathcal{R},\ \forall j \in \mathcal{S}_i,\ \forall m \in \mathcal{K}.
\end{equation}

Alternatively, when using quadratic least squares, the corresponding constraints are defined as follows. For cut sections, we impose for all $i$, $j$, $m$ with $u_{i,j,m} \leq 0$,
\begin{equation}
\mathcal{V}_{i,j,m}^{+} \geq \chi_{i,j,m,1} u_{i,j,m}^2 + \chi_{i,j,m,2} u_{i,j,m} + \chi_{i,j,m,3}.
\end{equation}
For fill sections, the constraint is given by for all $i$, $j$, $m$ with $0 \leq u_{i,j,m} \leq \bar{u}_{i,j,m}$,
\begin{equation}
\mathcal{V}_{i,j,m}^{-} \geq \chi_{i,j,m,1} u_{i,j,m}^2 + \chi_{i,j,m,2} u_{i,j,m} + \chi_{i,j,m,3}.
\end{equation}

A similar formulation is applied at intersection nodes. When using linear least squares, the cut condition for an intersection (\( z_{e,m} \leq 0 \)) is,
\begin{equation}
\mathcal{U}_{e,m}^{+} \geq \chi_{e,m,1} z_{e,m} + \chi_{e,m,2}, 
\quad \forall e \in \mathcal{I},\ \forall m \in \mathcal{K}.
\end{equation}
For fill (\( 0 \leq z_{e,m} \leq \bar{z}_{e,m} \)), we impose,
\begin{equation}
\mathcal{U}_{e,m}^{-} \geq \chi_{e,m,1} z_{e,m} + \chi_{e,m,2}, 
\quad \forall e \in \mathcal{I},\ \forall m \in \mathcal{K}.
\end{equation}

When using quadratic least squares at intersections, the constraint for cut becomes for all $e \in \mathcal{I}$, $m \in \mathcal{K}$,
\begin{equation}
\mathcal{U}_{e,m}^{+} \geq \chi_{e,m,1} z_{e,m}^2 + \chi_{e,m,2} z_{e,m} + \chi_{e,m,3}, 
\quad \text{for } z_{e,m} \leq 0.
\end{equation}
For fill, the corresponding constraint is,
\begin{equation}
\mathcal{U}_{e,m}^{-} \geq \chi_{e,m,1} z_{e,m}^2 + \chi_{e,m,2} z_{e,m} + \chi_{e,m,3}, 
\quad \text{for } 0 \leq z_{e,m} \leq \bar{z}_{e,m}.
\end{equation}

\bibliography{wileyNJD-APA}%

\end{document}